\documentstyle[preprint,floats,tighten,prl,aps]{revtex}
\def \pom {{\scriptscriptstyle \kern -0.1em I \kern -0.25em P}}
\def \lt {\mbox{\boldmath $\ell$}}

\def \MSbar {\vbox{\hrule\kern 1pt\hbox{\rm MS}}}

\input epsf.tex
\def\desepsf(#1 width #2){\epsfxsize=#2 \epsfbox{#1}}
\begin{document}
\preprint{\vbox{
\hbox{PSU-TH/224} 
\hbox{August 2000}
}  }
\draft

\title{ Color transparency  
in deeply inelastic  diffraction}
\author{F.\ Hautmann$^{a}$ and D.E.\ Soper$^{b}$}
\address{$^{a}$ Department of Physics, 
Pennsylvania State University, University Park PA 16802}
\address{$^{b}$ Institute of Theoretical Science, 
University of Oregon, Eugene OR 97403}

\maketitle

\begin{abstract}
We suggest a simple physical picture for the diffractive parton
distributions that appear in diffractive deeply inelastic scattering. In
this picture, partons impinging on the proton can have any transverse
separation, but only when the separation is small can they penetrate the
proton without breaking it up. By comparing the predictions from this
picture with the diffractive data from HERA, we determine rough values
for the small separations that dominate the diffraction process.
\end{abstract} 



Diffractive deeply inelastic scattering (DIS) has been extensively
studied at HERA \cite{abra}. For the theoretical description of this
process, one can usefully factor the perturbative short distance physics
from the non-perturbative long distance physics. In such a description,
the diffractive part is long distance and is described by diffractive
parton distributions.   The measured diffractive structure function
$F_2^{\rm diff}$ is  a convolution of  these  distributions with the
usual hard scattering  coefficients known from inclusive DIS. This is
described in detail in Ref.~\cite{hksnpb} and the references cited there. 

One of the results of Ref.~\cite{hksnpb} is that  the diffractive gluon
distribution,  although it contributes  to diffractive DIS only at the
next-to-leading order in $\alpha_s$, is extremely important in
diffractive DIS because it is  very large.  The magnitude of its ratio to
the quark distribution is found to be  governed by the color factor
$C_A^2 (N_c^2-1)/ (C_F^2 N_c) = 27/2$.  A very large gluon distribution 
makes diffraction an important  process to probe  the onset of  parton
saturation~\cite{muedis98,golecwue,levin} and the limits of 
applicability of perturbative evolution~\cite{ffss}.  Recent experimental
observations of  diffractive jet  production~\cite{osaka}, which  are
especially sensitive measurements of  the gluon distribution,   confirm
the existence of a very large diffractive gluon component.  In this
paper, we build on the  results of Ref.~\cite{hksnpb} to develop a 
physical picture for the diffractive gluon distribution  based on the
analysis of  absorption (color opacity) and scattering
(color transparency) effects in the diffraction process. 

The diffractive gluon distribution is  a product of matrix elements of
the gluon measurement operator
\begin{equation}
\widetilde G_a(0,y^-,{\bf 0})^{+j} = 
\left[{\cal P}\exp\left(-ig \int_{y^-}^\infty
\!dx^- A_c^+(0,y^-,{\bf 0}) t_c
\right)\right]_{ab}
G_b(0,y^-,{\bf 0})^{+j}\ 
, 
\label{opdef}
\end{equation}
where, in standard notation,  $G$ is the gluon field strength, $A$  the 
vector potential, $t$ the generators of the adjoint representation of 
$SU(3)$, and ${\cal P}$ denotes path ordering of the exponential.  
As explained in Ref.~\cite{hksnpb}, when viewed
in the proton rest frame~\cite{bjrest} this operator creates a state 
consisting of partons that have large momentum in the minus direction
($k^\pm = (k^0 \pm k^3)/\sqrt{2}$). These partons travel a long distance,
then impinge on the proton and scatter. The operator in Eq.~(\ref{opdef})
contains an eikonal line operator that is capable of absorbing gluons
from the color field of the proton and is equivalent to a color octet
parton moving with infinite minus-momentum.  Note the correspondence with
the  space-time description of diffractive DIS~\cite{niko,bart,buch}, in 
which a current operator $J^{\mu}(0,y^-,{\bf 0})$ creates a state of
large momentum partons that impinge on the proton and scatter. In fact,
in the diffractive gluon distribution, the color octet eikonal operator
is a stand-in for a quark-antiquark pair  that was produced by the
current but is too compact to be resolved by the color field of
the proton.

The partonic system created by the gluon operator can have any
transverse size when it reaches the proton.  We  propose here that,  if
the partons reach the proton at a large separation,  they cannot get
through the proton without breaking it up. Thus  diffractive deeply
inelastic scattering results mainly from the incoming partons being
close together and interacting with the color field of the proton only
through their color dipole moment. This would seem to make diffraction
quite unlikely, but the calculation shows that the partonic wave
function is large at small transverse separations between the
partons. Because of the short distance behavior of the
perturbative wave function, diffraction turns out to be not at all
unlikely.

In the following, we first restate the result~\cite{hksnpb} for the 
diffractive gluon distribution. Then we  formulate the  picture for the
interaction of the partonic system  created by the gluon operator with 
the proton's color field. We  use this in conjunction with the short
distance factorization of the  hard scattering   to calculate  the
diffractive structure function $F_2^{\rm diff}$. Finally we compare the
results to experiment and  extract a rough value for how close together
the partons have to be in order to make it through the proton without
breaking it apart.

In the notation of Fig.~1, the expression for the diffractive gluon
distribution is
\begin{equation}
{{d\, f_{g/A}^{\rm {diff}}(\beta , x_\pom , t , \mu ) } \over
{dx_\pom\,dt}} = 
{1 \over {16 \, \pi^3}} \, 
{N_c^2 -1 \over \beta (1-\beta) x_\pom^2 }
{1 \over 2}\sum_{s_{\!A},s_{\!A^\prime},s}
\int{ d^2{\bf k} \over (2\pi)^2}\
|{\cal G}|^2 
\label{diffpdffromG}
\end{equation}
with
\begin{equation}
{\cal G}^{j s}=
{ 1 \over N_c^2 - 1}\
\int{ d p_{\!A^\prime}^+ \over (2\pi) 2 p_A^+}\
\int dy^- e^{i\beta x_\pom p_{\!A}^+ y^-}
\langle p_{\!A^\prime},s_{\!A^\prime}; k,s,a|
\widetilde G_a(0,y^-,{\bf 0})^{+j}| p_{\!A},s_{\!A} \rangle 
\hspace*{0.2 cm} . 
\label{calGdef}
\end{equation}
The expression in Eqs.~(\ref{diffpdffromG}) and (\ref{calGdef})
represents a perturbative approximation in which the high momentum
gluon created by the operator $\widetilde G$ travels to the final state
without splitting into any further high momentum partons, so that there
is a single final state gluon with momentum $k$ in addition to the final
state proton. If one worked to higher orders of perturbation theory,
then there could be partonic splitting and joining and the final state
could contain multiple high momentum partons.  We integrate over the
transverse momentum ${\bf k}$ of the final state gluon; its plus
momentum is set to $k^+ = (1 - \beta) x_\pom p_A^+$ so that the total
plus-momentum removed from the proton is $x_\pom p_A^+$.   Its color $b$
must be $b = a$ since the initial and diffracted proton states are
color singlets.

\begin{figure}[htb]
\centerline{\desepsf(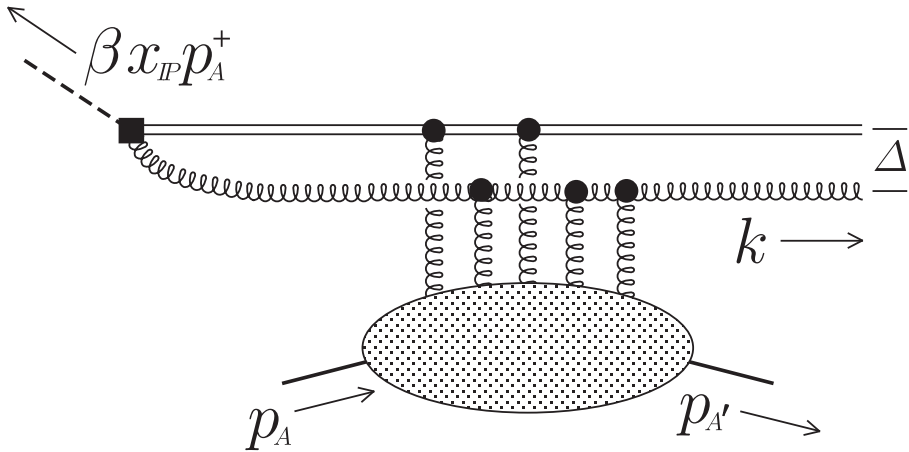 width 12 cm)}
\vspace*{15mm}
\caption{Structure of the diffractive gluon distribution. The top 
part is treated perturbatively. Any number of soft gluons are exchanged 
with the hadron matrix element represented on the bottom.}
\label{firstfig}
\end{figure}

At large $1 / x_\pom$ an eikonal approximation applies. There is a wave
function $\psi$, describing the state of the fast moving gluon as it
reaches the proton. This gluon passes through the proton at transverse
position ${\bf \Delta}$ while the eikonal operator in Eq.~(\ref{opdef})
is located at transverse position ${\bf 0}$. The result, taken after a
little manipulation from Eq.~(3.38) of Ref.~\cite{hksnpb}, is
\begin{eqnarray}
\label{Gfrom338}
{\cal G}^{j s} &\approx&
{ -1 \over N_c^2 - 1}\
\int{ d p_{\!A^\prime}^+ \over (2\pi) 2 p_A^+}\ 
\int {d^2{\bf \Delta}\over{\bf \Delta}^2}\, 
\int{ d^2 \lt \over (2\pi)^2} \,  
e^{- i \lt \cdot {\bf \Delta}}\,
{\partial  \over \partial \ell^i}{\partial  \over \partial \ell^i}
\psi^{js}(\beta, {\bf k},{\bf k}+\lt)
\nonumber\\
&&\times
\langle p_{\!A^\prime},s_{\!A^\prime}|
{\rm Tr}\left[1 - F({\bf \Delta}) F^T({\bf 0})\right]
| p_{\!A},s_{\!A} \rangle \hspace*{0.2 cm} .
\end{eqnarray}
Here the operators $F$  are color octet eikonal line operators:
\begin{equation}
F({\bf z}) = 
{\cal P}\ \exp\left\{
-ig\int_{-\infty}^{+\infty} d z^- {\cal A}_a^+(0,z^-,{\bf z})\, t_a
\right\} \hspace*{0.2 cm} .  
\label{Fdef}
\end{equation} 
The wave function for the gluon state is
\begin{equation}
\psi^{js}(\beta,{\bf k},{\bf p}) = 
{ \beta {\bf k}^2 \delta^{js} + 2(1-\beta) p^j p^s
\over \beta {\bf k}^2 + (1-\beta) {\bf p}^2}.
\end{equation}
We choose here to differentiate $\psi$ twice with respect to $\lt$. This
has the effect of eliminating apparent ultraviolet divergences from the
integration over $\bf k$ in the formulas that follow. The derivatives
with respect to $\lt$ are compensated by a factor $(-1/{\bf\Delta}^2)$
multiplying the matrix element of the eikonal operators. The factor
$1/{\bf\Delta}^2$ does not itself produce a short distance divergence
because both the matrix element and its first derivative with respect to 
${\bf\Delta}$ vanish at ${\bf\Delta} = {\bf 0}$. See Sec.~4.1 of
Ref.~\cite{hksnpb} for a discussion of the ultraviolet behavior in the
minimal case of two gluon exchange; exchanging more gluons does not make
the ultraviolet behavior worse.

Consider now the matrix element of the eikonal operators between
the initial and final proton states. The initial state carries
zero transverse momentum. Let $-{\bf q}$ be the transverse momentum 
carried by the final state (${\bf q}^2 \approx -t$).  
We may first expand this state in terms of 
eigenstates of transverse position with transverse position $-{\bf b}$ as
\begin{equation}
\label{expandb}
\langle p_{\!A'},s_{\!A'}| = \langle p_{\!A'}^+,-{\bf q},s_{\!A'}|
= \int d^2{\bf b}\ e^{-i\,{\bf b}\cdot {\bf q}}\
{}_x\!\langle p_{\!A'}^+,-{\bf b},s_{A'}| \hspace*{0.2 cm} , 
\end{equation}
where the subscript $x$ on a state vector denotes  that it is a position
eigenstate.  We then apply a  translation to put the proton at
transverse position ${\bf 0}$ and let the eikonal operators be at
transverse positions ${\bf b}$ and ${\bf b}  + {\bf \Delta}$. Taking
into account that the eikonal operators do not transfer any plus
momentum, we define a non-perturbative function $\Xi$ as 
\begin{eqnarray}
\label{defxi}
&& (2\pi) \ 2 p_{\!A}^+ \ \delta(p_{\!A'}^+ - p_{\!A}^+)\
\Xi({\bf b} + {\bf \Delta},{\bf b};s_{\!A'},s_{\!A})
\nonumber\\ 
&& \hspace*{0.3 cm} = 
{1 \over {N_c^2 - 1}} \, 
{}_x\!\langle p_{\!A'}^+,{\bf 0},s_{\!A'}|
{\rm Tr}\left[1 - F({\bf b} + {\bf \Delta}) F^T({\bf b})\right]
|p_{\!A}^+,{\bf 0},s_{\!A}\rangle
 \hspace*{0.2 cm} .
\end{eqnarray}

We use Eqs.~(\ref{expandb}) and (\ref{defxi}) in
Eqs.~(\ref{diffpdffromG}) and (\ref{Gfrom338}) to give
\begin{eqnarray}
\label{arrive}
\lefteqn{{{d\, f_{g/A}^{\rm {diff}}(\beta , x_\pom , t , \mu ) }
\over {dx_\pom\,dt}} 
= 
\int 
{d^2{\bf \Delta}\over {\bf \Delta}^2}\, 
{d^2{\bf \Delta}'\over {{\bf \Delta}'}^2}\,
d^2{\bf b}\,d^2{\bf b}'\,
e^{-i\,({\bf b} - {\bf b}')\cdot {\bf q}}\
{\cal I}(x_\pom,\beta,{\bf \Delta},{\bf \Delta}')}
\nonumber\\
&&\times
{1 \over 2}\sum_{s_{\!A},s_{\!A^\prime}}
\Xi^*({\bf b}' + {\bf \Delta}',{\bf b}';s_{\!A'},s_{\!A})\,
\Xi({\bf b} + {\bf \Delta},{\bf b};s_{\!A'},s_{\!A})
\end{eqnarray}
where
\begin{eqnarray}
\label{Ifun}
&& {\cal I}(x_\pom,\beta,{\bf \Delta},{\bf \Delta}') =
{1 \over {16 \, \pi^3}} \, 
{N_c^2 -1 \over \beta (1-\beta) x_\pom^2 }
\int{ d^2{\bf k} \over (2\pi)^2}\,
\int { d^2 {\lt} \over (2\pi)^2}\,
 e^{- i {\lt}\cdot{\bf \Delta}} \int 
{ d^2{\lt}' \over (2\pi)^2}\
e^{i {\lt}'\cdot{\bf \Delta}'}\
\nonumber\\
&&\times 
\left[
{\partial  \over \partial {\ell'}^k}
{\partial  \over \partial {\ell'}^k}\,
 \psi^{js}(\beta,{\bf k},{\bf k} +\lt') 
\right]^*
\, \left[ 
{\partial  \over \partial \ell^i}{\partial  \over \partial \ell^i}\,
\psi^{js}(\beta,{\bf k},{\bf k} + \lt) 
\right] 
\hspace*{0.2 cm}  .
\end{eqnarray}

Eq.~(\ref{arrive}) provides a representation for the diffractive gluon
distribution in a physically appealing form.  The $\beta$ dependence is
entirely contained in the factor ${\cal I}$, which is constructed from
the perturbative wave functions $\psi$. There are integrals over  the
global impact parameter of the collision (${\bf b}$ for the amplitude
and ${\bf b}'$ for the complex conjugate). The factor $\cal
I$ does not depend on ${\bf b}$ or ${\bf b}'$; these parameters merely
tell where the proton is relative to the incoming partons. There are
also integrals over the parton transverse separation (${\bf \Delta}$
for the amplitude  and ${\bf \Delta}'$ for its complex conjugate). The
perturbative function ${\cal I}/({\bf \Delta}^2\,{{\bf \Delta}'}^2)$
gives the amplitudes for the partons to have separations ${\bf \Delta}$
and ${\bf \Delta}'$; then the functions $\Xi$ describe the effect of the
proton's soft color field.

If we integrate the diffractive gluon distribution (\ref{arrive}) 
over $x_\pom$ at fixed $x$ (or equivalently over $\beta$),    
 this integration 
produces a logarithmic divergence coming
from the region $\beta = x/x_\pom \ll 1$.
In the case of the quark distribution there is 
no such divergence~\cite{hksnpb}. Then by integrating over $\beta$ 
one obtains a formula with only one separation 
parameter ${\bf \Delta}$. Formulas of this kind are 
given, e.g., in Refs.~\cite{golecwue,niko,buch}. 
For the  discussion 
that follows we are interested in  the full $\beta$ spectrum, 
for which formulas with  two separation parameters, 
${\bf \Delta}$ and ${\bf \Delta}'$, apply.

If we integrate the diffractive gluon distribution (\ref{arrive})
over $t = -{\bf q}^2$ we obtain
\begin{eqnarray}
\label{arrivebis}
\lefteqn{{{d\, f_{g/A}^{\rm {diff}}(\beta , x_\pom ,  \mu ) }
\over {dx_\pom}} 
= 
4\pi
\int 
{d^2{\bf \Delta}\over {\bf \Delta}^2}\, 
{d^2{\bf \Delta}'\over {{\bf \Delta}'}^2}\,
d^2{\bf b}\
{\cal I}(x_\pom,\beta,{\bf \Delta},{\bf \Delta}')}
\nonumber\\
&&\times
{1 \over 2}\sum_{s_{\!A},s_{\!A^\prime}}
\Xi^*({\bf b} + {\bf \Delta}',{\bf b};s_{\!A'},s_{\!A})\,
\Xi({\bf b} + {\bf \Delta},{\bf b};s_{\!A'},s_{\!A}).
\end{eqnarray}
Now there is only one impact parameter, $\bf b$, but there are still
two parton separation variables, ${\bf \Delta}$ and ${\bf \Delta}'$.

We now turn to the hadronic matrix element $\Xi$. If, as in
Ref.~\cite{hkslett}, the proton were replaced by a small meson made of
heavy quarks, then perturbation theory would be applicable to compute 
$\Xi$. But for a large state, such as a proton, we need to rely on a
combination of perturbative and nonperturbative ideas to construct an
approximation for $\Xi$.

We first recognize that $\Xi$ is an absorption amplitude. Imagine that
two high momentum gluons, represented by eikonal operators, impinge upon
the proton at transverse positions ${\bf b}$ and ${\bf b} + {\bf
\Delta}$.  Suppose that the gluons completely miss the proton. Then
each of the two operators $F$  in $1 - F({\bf \Delta})F^T({\bf 0})$ can be
replaced by unit operators so that $\Xi = 0$. Now, suppose
that one or both of ${\bf b}$ and ${\bf b} + {\bf \Delta}$ is inside the
proton, with  no particular relation between these positions. Then we
expect that the amplitude for the proton to remain a ground state proton
after the passage of the gluon(s) is approximately zero, $\langle
p|F\,F^T|p \rangle \approx 0$, so that $\Xi \approx 1 \times \delta_{s_A
s_{A'}}$. Thus a simple model for $\Xi $ might be
\begin{equation}
{\cal A}({\bf b} + {\bf \Delta},{\bf b})\
\delta_{s_A s_{A'}}
\end{equation}
where $\cal A$ is a smooth function such that ${\cal A} \approx 0$ if 
$|{\bf b} + {\bf \Delta}| \gg R_p^2$ and $|{\bf b}| \gg R_p^2$ and
${\cal A} \approx 1$ if either $|{\bf b} + {\bf \Delta}| \ll R_p^2$ or
$|{\bf b}| \ll R_p^2$, where $R_p$ is the radius of the proton.

An important exception arises if ${\bf \Delta}$ is small.
Then $F({\bf b})^T$ is nearly the inverse of $F({\bf b} + {\bf \Delta})$,
so that $\Xi \approx 0$. To take this color transparency effect into
account, we modify our model to
\begin{equation}
{\cal A}({\bf b} + {\bf \Delta},{\bf b})\
\delta_{s_A s_{A'}}\,
[1 - \exp(- \kappa_g^2{\bf \Delta}^2/2)].
\label{gaussian}
\end{equation}
Then $\Xi \approx 0$ for small $\bf \Delta$ but for large $\bf
\Delta$ we retain the absorption physics explained above. Here
$1/\kappa_g$ is a parameter that measures how close together the gluons
have to be before they can traverse the proton without breaking it up.
That is, $1/\kappa_g$ is the largest size for a color octet dipole for
which the proton is transparent. We hypothesize that $1/\kappa_g$ is
small:
$
1/\kappa_g \ll R_p.
$

In Eq.~(\ref{gaussian}) we have chosen a gaussian function for reasons
of simplicity. Any function that vanishes for $\kappa_g {\bf \Delta} \ll
1$ and tends to 1 for $\kappa_g {\bf \Delta} \gg 1$ could be used in
place of  $[1 -
\exp(- \kappa_g^2{\bf \Delta}^2/2)]$.

We now note that the function $\cal I$ is singular when ${\bf \Delta}$
and ${\bf \Delta}'$ become small. In fact, straightforward dimensional
analysis shows that for $|{\bf \Delta}| \sim |{\bf \Delta}'|$, 
${\cal I} \sim {\mbox{const.}}/(|{\bf \Delta}|\, |{\bf \Delta}'|)$. 
For this reason,
the important contributions to the integral over ${\bf \Delta}$ comes
from small ${\bf \Delta}$, $|{\bf \Delta}| \sim 1/\kappa_g \ll R_p$. In
this circumstance, we can set ${\bf\Delta}$ to zero in ${\cal A}({\bf b}
+ {\bf \Delta},{\bf b})$, giving
\begin{equation}
\label{Gmodel}
\Xi({\bf b} + {\bf \Delta},{\bf b};s_{\!A'},s_{\!A})
=
{\cal A}({\bf b},{\bf b})\,
\delta_{s_A s_{A'}}\,
[1 - \exp(- \kappa_g^2{\bf \Delta}^2/2)].
\end{equation}

This is our model. The proton can survive, $\Xi \approx 0$, if either
${\bf b}$ is large or ${\bf \Delta}$ is small. 
The model incorporates 
saturation~\cite{muedis98,golecwue,buch}:  
as $|{\bf \Delta}|$ increases, the absorption amplitude
increases, but it never gets to be larger than 1. Furthermore, the 
important transverse
separations are  small~\cite{hksnpb,muedis98,golecwue,buch}.

We now use Eq.~(\ref{Gmodel}) in Eq.~(\ref{arrivebis}).
The integral in ${\bf b}$ over the absorption function $\cal A$ 
must give a quantity of order $R_p^2$:  
\begin{equation}
\label{setintb}
\int d^2{\bf b} \  
|{\cal A}({\bf b},{\bf b})|^2
 = \lambda_{\cal A} \ R_p^2 \hspace*{0.2 cm}  ,  
\end{equation} 
where the precise 
coefficient $\lambda_{\cal A}$ depends on the detailed form of 
the function $\cal A$.  In the numerical evaluations that follow we take
$R_p$ to be the charge radius of the proton, $R_p = 0.88\ {\rm fm}$
\cite{prorad}, and we take $\lambda_{\cal A} = 1$. Using
Eq.~(\ref{setintb}) and Eq.~(\ref{Ifun}) for ${\cal I}$ in
Eq.~(\ref{arrivebis}), we see that
\begin{equation}
\label{simplestr}  
{{d\, f_{g/A}^{\rm {diff}} }
\over {dx_\pom}} =  
{{(N_c^2 -1)\,  \lambda_{\cal A} \ R_p^2 \ \kappa_g^2} \over  x_\pom^2 } \
g(\beta) 
\hspace*{0.2 cm} ,
\end{equation}
where $g(\beta)$ is a function of $\beta$ that we  
determine by calculating the integrals numerically in a suitable
momentum space representation.  The behavior in $\kappa_g$ is perhaps
surprising -- the larger $\kappa_g$, the larger the cross section. But
it reflects an essential feature of the diffraction process: because of
the singular nature of the wave functions the amplitude is dominated by
distances of order $1 /  \kappa_g$. 

Notice that the result (\ref{simplestr}) has an $x_\pom$ dependence 
$x_\pom^{-2}$. This results from our simple treatment of the soft
exchanged gluons, which we treat as having rapidities near zero in
the proton rest frame. In a more sophisticated treatment, one would have a
BFKL ladder of exchanged gluons with rapidities becoming large and
negative along the ladder. This would lead to a dependence
$x_\pom^{-2- 2\delta}$ with $\delta$ proportional to $\alpha_s$. We do
not pursue such a treatment here. However, in order to compare with
experimental data, we suppose that the simple treatment is
approximately correct for $x_\pom$ of order $x_0$, where $x_0$ is small
but not too small, and that there are corrections for $x_\pom
\ll x_0$. Thus we multiply our result by $x_0^{2 \delta}/x^{2 \delta}$.
We use the experimental value $\delta = 0.15$ \cite{h1xp,zeusf2d2} and
take $x_0 = 0.01$. This gives

\begin{equation}
\label{simplestr2}  
{{d\, f_{g/A}^{\rm {diff}} }
\over {dx_\pom}} =  
{{(N_c^2 -1)\,  x_0^{2\delta}\, \lambda_{\cal A}\, \ R_p^2 \ 
\kappa_g^2} \over  x_\pom^{2 + 2 \delta} } \
g(\beta) 
\hspace*{0.2 cm} .
\end{equation}

A treatment analogous to that described for the gluon may now be
applied to the quark distribution, using the appropriate quark operators
as described in \cite{hksnpb}. Similarly to Eq.~(\ref{simplestr}),  the
answer has the form
\begin{equation}
\label{quarkstr}  
{{d\, f_{q/A}^{\rm {diff}} }
\over {dx_\pom}} =  
{{N_c  \, x_0^{2\delta}\,\lambda_{\cal A} \, R_p^2 \, \kappa_q^2} 
\over  x_\pom^{2 + 2 \delta} } \
q(\beta) 
\hspace*{0.2 cm} .
\end{equation}
The function $q(\beta)$ is different than in the gluon case due to the
different lightcone wave functions $\psi$.   The color factor is
different. Finally, the  scale $\kappa_q$ is different because  color
triplet eikonal operators now replace color octet operators in the
functions $\Xi$.  From the color structure of these operators, we expect
$\kappa_g$ and $\kappa_q$ to be  roughly in the relation $ \kappa_g
/\kappa_q \approx C_A / C_F$.  Note that, together with the color factors
explicitly  shown in Eqs.~(\ref{simplestr2}) and (\ref{quarkstr}),  this
gives back the factor  $C_A^2 (N_c^2-1) / (C_F^2 N_c) = 27/2$ of
Ref.~\cite{hksnpb} in the ratio between the gluon and the quark
distributions. Note also that, since $\kappa_q < \kappa_g$, the typical
separation ${\bf \Delta}$ in the case of quarks will be larger than in the
case of gluons, so that the perturbative evaluation that we have used may
be less reliable in the quark case.

The treatment discussed so far does not contain scaling violation. 
Having worked in a short-distance factorization framework, we may include
the graphs that give scaling violation systematically by using
next-to-leading-order hard scattering coefficients and  DGLAP evolution
equations, as in Ref.~\cite{hksnpb}. We then compute predictions for the
diffractive structure function $F_2^{\rm {diff}}$.

What, then, should be the starting scale $\mu_0$ for evolution?
Evidently, $\mu_0$ should be roughly comparable to the physical scales
$\kappa_g$ and $\kappa_q$. To go further, consider the term $P_{qg}
\otimes f_g$ in the evolution equation for the diffractive quark
distribution at scale $\mu_0$. The quark and antiquark have a
transverse separation of order $1/\mu_0$. The quark-antiquark system is
separated from the gluon by a larger distance, which, for us, is of order
$1/\kappa_g$. Thus $1/\mu_0$ should be somewhat smaller than
$1/\kappa_g$.

\begin{figure}[htb]
\vspace{115mm}
\includegraphics{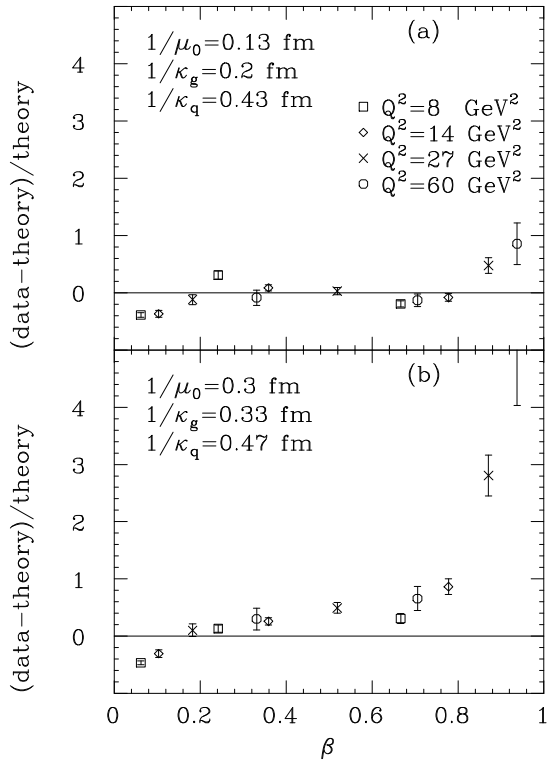}
\caption{The data for $F_2^{\rm {diff}}$ 
from Ref.~\protect\cite{zeusf2d2} compared with the theory 
results described in the text for two different sets of 
scale parameters.}
\label{secondfig}
\end{figure}

Our results for $F_2^{\rm {diff}}$ depend on $\kappa_g$, $\kappa_q$, and
$\mu_0$. Approximate values for these scales may be extracted by
comparing the predictions with experiment. Note that a very precise
determination is difficult because of the uncertainty in the
normalization of the model, which arises from the uncertainty in the
factors $\lambda_{\cal A}$ and $x_0^{2\delta}$ in Eqs.~(\ref{simplestr2})
and (\ref{quarkstr}). This normalization uncertainty affects the
determination of $\kappa_q$ directly. The parameter $\mu_0$ and the ratio
$\kappa_q/\kappa_g$ influence the shape of $F_2^{\rm diff}$ in $\beta$
and $Q^2$, so their determination is less directly affected.
 
In Fig.~2  we  compare the results for $F_2^{\rm {diff}}$  with the
data of Ref.~\cite{zeusf2d2} by plotting data minus theory  over 
theory, where the theory in the two plots is the one  corresponding to
the two displayed sets of values for $1/ \mu_0$, $1/ \kappa_g$, $1/
\kappa_q$.  Recall that, for the self-consistency  of the physical
picture that we have discussed,  we expect these distance scales to be
small compared to $R_p$.  For the case of Fig.~2a, we find that the
agreement between theory and data is reasonable, given the  simplicity
of the model. The size $1 / \mu_0$ of the quark-antiquark pair here is 
smallest  and the ratio $\kappa_g / \kappa_q$ of the sizes of the
color-triplet to the color-octet  systems comes out to be in rough
agreement with the expectation $C_A / C_F$   mentioned above. In
Fig.~2b we increase the value of $1 / \mu_0$ and then adjust  $1 /
\kappa_g$ and $1 / \kappa_q$ so as to describe the data as well as
possible.  We observe that the data seem to disfavor these larger
distance scales:   the shape of the data, in particular, is not well
reproduced in this case.  

It is now worth looking back at Fig.~1 and the structure of 
Eq.~(\ref{arrivebis}). 
 The function $\Xi$, representing the lower part of Fig.~1 and    
 including the exchange of any number of soft 
gluons, is nonperturbative. We have posited a  description 
for  $\Xi$ in terms of an absorption function and a color transparency 
factor. 
The   upper part of Fig.~1  
is also, in general,  
nonperturbative, because it still contains arbitrarily large   
parton separations $|{\bf \Delta}|$. 
Indeed,  in the  upper part of Fig.~1 
one would expect to have
not just two but several fast partons impinging on the proton, each with a
large transverse separation with respect to the others. But 
  if the separations are large, then 
absorption  dominates the  hadronic matrix element, so that 
central collisions are unlikely to  
leave the proton intact. Thus, the parton transverse 
separations  
 must be smaller than  the color transparency scale. 
If this  scale is  small enough,  
 a perturbation expansion  
for the upper part of Fig.~1 is justified. At
lowest perturbative order there are only two partons (including the
eikonal line) and we obtain the calculated function $\cal I$. At higher 
perturbative orders there can be multiple partons, all close together;  
these are the configurations that we include through  
radiative corrections to the hard 
scattering and evolution.

The results of Fig.~2 
provide a justification for this  expansion and 
lend support to the physical picture of
diffractive DIS as being due to color-dipole parton systems penetrating
the proton at small transverse separations. In particular, a partonic
system in a configuration of a color octet dipole, which corresponds to
the diffractive gluon distribution, must be especially small in
transverse size. This contributes to the  diffractive gluon
distribution being especially large, as is consistent with the data.
The picture described in this paper provides a concrete model for the
mechanism through which the dominance of small size
states~\cite{hksnpb,hkslett}  may be realized in diffractive DIS at
HERA.  More and better data will be valuable to investigate this
further and decide whether or not the indications of Fig.~2 are
confirmed. 

\vskip 0.3 cm

We gratefully acknowledge discussions with J.~Collins, M.~Diehl, 
Z.~Kunszt, A.~Mueller, G.~Sterman and M.~Strikman.  Our work is 
supported in part by the US Department of Energy under 
grants DE-FG02-90ER40577 and DE-FG03-96ER40969. 
F.H. acknowledges the hospitality of  the University of 
Oregon  while part of this work was being done.

\end{document}